\documentclass[preprint,showpacs,preprintnumbers,amsmath,amssymb]{revtex4}

\usepackage{graphicx}
\usepackage{dcolumn}
\usepackage{bm}

\begin{document}


\title{Phase transitions in LaFeAsO: structural, magnetic, elastic, and transport properties, heat capacity and M\"{o}ssbauer spectra}

\author{Michael A. McGuire}
\affiliation{Oak Ridge National Laboratory, Oak Ridge, Tennessee 37831 USA}
\author{Andrew D. Christianson}
\affiliation{Oak Ridge National Laboratory, Oak Ridge, Tennessee 37831 USA}
\author{Athena S. Sefat}
\affiliation{Oak Ridge National Laboratory, Oak Ridge, Tennessee 37831 USA}
\author{Brian C. Sales}
\affiliation{Oak Ridge National Laboratory, Oak Ridge, Tennessee 37831 USA}
\author{Mark D. Lumsden}
\affiliation{Oak Ridge National Laboratory, Oak Ridge, Tennessee 37831 USA}
\author{Rongying Jin}
\affiliation{Oak Ridge National Laboratory, Oak Ridge, Tennessee 37831 USA}
\author{E. Andrew Payzant}
\affiliation{Oak Ridge National Laboratory, Oak Ridge, Tennessee 37831 USA}
\author{David Mandrus}
\affiliation{Oak Ridge National Laboratory, Oak Ridge, Tennessee 37831 USA}
\author{Yanbing Luan}
\affiliation{Department of Materials Science and Engineering, The University of Tennessee, Knoxville, Tennessee 37996-2200 USA}
\author{Veerle Keppens}
\affiliation{Department of Materials Science and Engineering, The University of Tennessee, Knoxville, Tennessee 37996-2200 USA}
\author{Vijayalaksmi Varadarajan}
\affiliation{Department of Physics and Astronomy, University of Kentucky, Lexington, KY 40506-0055 USA}
\author{Joseph W. Brill}
\affiliation{Department of Physics and Astronomy, University of Kentucky, Lexington, KY 40506-0055 USA}
\author{Rapha\"{e}l P. Hermann}
\affiliation{Institut f\"ur Festk\"orperforschung, Forschungszentrum J\"ulich GmbH, D-52425 J\"ulich, Germany}
\affiliation{Department of Physics, B5, Universit\'e de Li\`ege, B-4000 Sart-Tilman, Belgium}
\author{Moulay T. Sougrati}
\affiliation{Department of Physics, B5, Universit\'e de Li\`ege, B-4000 Sart-Tilman, Belgium}
\author{Fernande Grandjean}
\affiliation{Department of Physics, B5, Universit\'e de Li\`ege, B-4000 Sart-Tilman, Belgium}
\author{Gary J. Long}
\address{Department of Chemistry, Missouri University of Science and Technology, Rolla, MO 65409-0010, USA}

\date{\today}

\begin{abstract}
We present results from a detailed experimental investigation of LaFeAsO, the parent material in the series of ``FeAs'' based oxypnictide superconductors. Upon cooling this material undergoes a tetragonal-orthorhombic crystallographic phase transition at $\sim$160 K followed closely by an antiferromagnetic ordering near 145 K. Analysis of these phase transitions using temperature dependent powder X-ray and neutron diffraction measurements is presented. A magnetic moment of $\sim$0.35 $\mu_B$ per iron is derived from M\"{o}ssbauer spectra in the low temperature phase. Evidence of the structural transition is observed at temperatures well above the structural transition (up to near 200 K) in the diffraction data as well as the polycrystalline elastic moduli probed by resonant ultrasound spectroscopy measurements. The effects of the two phase transitions on the transport properties (resistivity, thermal conductivity, Seebeck coefficient, Hall coefficient), heat capacity, and magnetization of LaFeAsO are also reported, including a dramatic increase in the magnitude of the Hall coefficient below 160 K. The results suggest that the structural distortion leads to a localization of carriers on Fe, producing small local magnetic moments which subsequently order antiferromagnetically upon further cooling. Evidence of strong electron-phonon interactions in the high-temperature tetragonal phase is also observed.
\end{abstract}

\maketitle

\section{Introduction}

The family of lanthanide iron oxypnictides crystallizing in the ZrCuSiAs structure type has received extensive attention in the recent condensed matter literature. Many of these compounds have been known for almost a decade \cite{Jeitschko2000, zimmer}, and superconductivity below $\sim$7 K was reported in fluorine doped LaMPO (M = Fe, Ni) in 2006 and 2007 \cite{Kamihara2006, Liang2007, Watanabe2007}. However, the discovery of superconductivity at 28 K in fluorine doped LaFeAsO \cite{Kamihara2008}, and the subsequent reports of transition temperatures greater than 50 K in some of the related rare-earth materials has generated great interest and a flurry of recent experimental and theoretical activity, generating preprints daily at arxiv.org/archive/cond-mat and numerous publications in just the first few months \cite{pub-Hunte, pub-Sefat, pub-Chen, SinghArXiv, pub-Sun, pub-delaCruz, pub-Wu, pub-Cheng, pub-Ishibashi, pub-Day, pub-Takahashi, pub-Han, pub-Haule, pub-Wen, pub-Hand, kito, sato, ishibashi, pub-Chen2, Ren, Ren2, Xu}. The superconductivity in these materials appears to be unconventional \cite{MazinArXiv, pub-Hunte, pub-Sefat}, and much careful work will be required to elucidate the interesting physics in this family of compounds. In the arsenic containing materials doping is required to produce the superconducting state, and much work is duely focussed on the doping behavior of these compounds. However, study of the undoped materials is also important in developing an understanding of the superconductivity.

The undoped material LaFeAsO has been reported to undergo a spin density wave (SDW) transition near 150 K \cite{Kamihara2008, DongArXiv}, based on specific heat, resistivity, and reflectivity measurements. A structural phase transition has also been reported in this material at temperatures just above the magnetic transition \cite{pub-delaCruz, Nomura}. Upon doping with fluorine the SDW is suppressed and superconductivity emerges \cite{DongArXiv, pub-Sefat, Kamihara2008}. Careful characterization of the behavior of LaFeAsO and other materials in this family is important in understanding the underlying physics responsible for these behaviors. Here we report the results of our experimental investigation of LaFeAsO. We present structural analysis through the crystallographic phase transition from powder X-ray diffraction (PXRD) and investigation of the magnetic transition using neutron powder diffraction (NPD) along with magnetization and M\"{o}ssbauer spectral measurements. Signatures of these phase transitions are observed in ac-calorimetry heat capacity and resonant ultrasound spectroscopy (RUS) measurements. The effects of the structural and magnetic transitions on the transport properties of LaFeAsO (electrical resistivity, magnetoresistance, Hall effect, Seebeck effect, and thermal conductivity) are also examined.

\section{Experimental Details}
\subsection{Sample synthesis}
We tried several methods for synthesizing high quality samples of LaFeAsO. Starting materials were from Alfa Aesar and included La (99.9 \%), La$_2$O$_3$ (99.99 \%), Fe (99.998 \%), Fe$_2$O$_3$ (99.998 \%), and As (99.999 \%). The binaries LaAs and FeAs were synthesized for use as precursors by heating \textit{slowly} over the course of several days stoichiometric mixtures of the elements in sealed, evacuated silica tubes to 950 $^{\circ}$C for LaAs and 1050 $^{\circ}$C for FeAs, and soaking for 12$-$24 hours. Different synthesis routes will be distinguished by capital letters A-D in the discussion below.

Synthesis route A used a finely ground stoichiometric mixture of LaAs, Fe, and Fe$_2$O$_3$ pressed into a pellet, wrapped in Ta foil and heated in a sealed silica tube partially backfilled with ultra-high purity Ar and with a small piece of Zr foil, and heated at 1200 $^{\circ}$C for about one day. Powder X-ray diffraction (PXRD) analysis showed the product to be nearly single phase LaFeAsO. The only other phases observed by PXRD were Fe and La$_{2}$O$_{3}$, while neutron diffraction experiments detected Fe$_2$As. Rietveld analysis \cite{Fullprof, expgui, gsas} indicates that the amount of each of these impurities was less than 5 \%. It is likely that the iron containing impurities result from the reaction of the Zr foil with As vapor in the tube, reducing FeAs (a typical impurity phase when no metal foils are present) to Fe$_2$As and Fe.

Synthesis route B is identical to A except no metal foils were used. This typically gives a product with significant ($\sim$10\%) La$_{2}$O$_{3}$ and FeAs impurities as determined by neutron diffraction and M\"{o}ssbauer spectroscopy, but no Fe nor Fe$_2$As.

Synthesis route C involves reacting the product of route B with a small excess of La powder (a few percent). We found that this helps to further react the La$_2$O$_3$ and FeAs impurities left from route B and can produce PXRD pure material. Occasionally small amounts of LaAs and/or La$_2$O$_3$ are observed in the products from route C.

Route D used a finely ground stoichiometric mixture of FeAs, La$_{2}$O$_{3}$, and La pressed into a pellet, sealed in a silica tube partially backfilled with ultra-high purity Ar, and heated at 1200$^{\circ}$C for 30-36 hours. This method produced purer samples, sometimes with no impurities observable by PXRD.

A single sample prepared by route A was used for neutron diffraction, Hall effect, resistivity, magnetoresistance, thermal conductivity, Seebeck coefficient, and heat capacity measurements reported below. The presence of elemental Fe precluded magnetic characterization of this sample; the material was strongly magnetic at room temperature. Magnetization measurements reported below were carried out on a PXRD pure sample obtained via route C. Material produced by route B was used for M\"{o}ssbauer spectroscopy and elastic constant measurements. The temperature dependent PXRD data presented here were collected using a sample made by route D, and are in good agreement with similar measurements on the sample used for transport, NPD, and heat capacity measurements (route A).

\subsection{Characterization techniques}
Temperature dependent PXRD data were collected using an Anton Parr TTK450 low temperature stage on a PANalytical X'pert Pro MPD with an X'celerator position sensitive detector and using Cu K$\alpha$ radiation. The PXRD sample was mixed with copper powder in a small amount of vacuum grease to calibrate the sample temperature for each scan using the refined lattice constant of Cu, and to increase the thermal contact between the sample and the holder.  Neutron diffraction measurements were performed at the High Flux Isotope Reactor (HFIR) on $\sim$1.5 g samples of LaFeAsO and LaFeAsO$_{0.89}$F$_{0.11}$ on the HB1A triple-axis spectrometer with horizonal collimations of 48-40-40-68 using a highly filtered beam ($\lambda$/2 $\sim$ $10^{-4}\lambda$) with an incident energy of 14.64 meV.  Preliminary measurements were also performed on the WAND diffractometer and the HB3 triple-axis spectrometer.  Data were collected over a range of scattering angles from 5$^{\circ}$ to 130$^{\circ}$ at several temperatures.

Transport measurements were performed using a Quantum Design Physical Property Measurement System (PPMS). Silver epoxy (Epotek H20E) was used for electrical and thermal contacts. Hall effect in fields up to 6 T and electrical resistivity measurements were made using platinum wire leads. Gold-coated copper leads were used for thermal conductivity and Seebeck coefficient measurements.

Heat capacity was determined by ac-calorimetry measurements made on a 3.18 mg polycrystalline sample, using 4.5 Hz chopped white light as a heating source \cite{accal}. The technique only yields relative values of the heat capacity, so the results were normalized at 127 K to lower resolution specific heat data taken with the PPMS. The data were then corrected for the addendum heat capacity (thermocouple wire plus glue), which was $\sim$ 2.5 \% of the sample's in the temperature region measured.

The elastic moduli were measured using Resonant Ultrasound Spectroscopy (RUS) and a custom made probe in a Quantum Design PPMS cryostat \cite{russetup}. Stanford Research synthesized function generator and Model SR 844 RF Lock-in Amplifier were used to excite the sample and collect the data.

The iron-57 M\"ossbauer spectra were recorded between 4.2 and 295 K on a constant acceleration spectrometer which utilized a rhodium matrix cobalt-57 source and was calibrated at 295 K with $\alpha$-iron powder. The isomer shifts are relative to room temperature $\alpha$-iron. The accuracy of the temperature is better than $\pm$1 \%.

\begin{figure}
\includegraphics[width=3.0in]{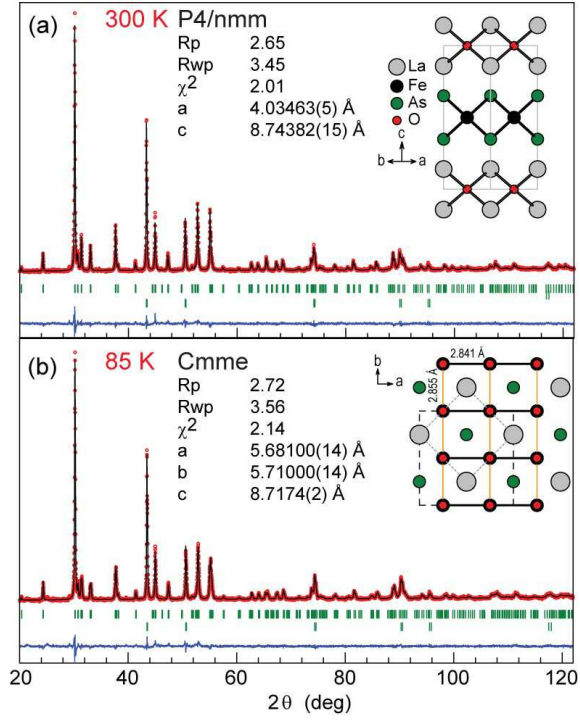}
\caption{\label{fig:pxrd}
(color online) Rietveld refinement results for LaFeAsO at (a) 300 K and (b) 85 K. The upper sets of ticks mark the location of Bragg peaks from LaFeAsO, while the lower ticks locate Bragg peaks from the Cu internal standard. The inset in (a) shows the tetragonal structure viewed along (1 1 0) emphasizing the FeAs layer in this material. The grey line outlines the tetragonal unit cell. The inset in (b) shows the orthorhombic structure viewed along (0 0 1), emphasizing the distortion which occurs in the ab-plane upon cooling through the tetragonal to orthorhombic phase transition. The dashed black lines outline the c-centered orthorhombic unit cell, while the dotted grey lines show the primitive unit cell (see text).
}
\end{figure}

\section{Results and discussion}

\subsection{X-ray diffraction: structural phase transition}
It has been reported that LaFeAsO undergoes a structural phase transition from tetragonal (P4/nmm, space group number 129) to orthorhombic (Cmme, space group formerly known as Cmma, number 67) upon cooling at about 155 K \cite{pub-delaCruz, Nomura}. Figure \ref{fig:pxrd} shows the Rietveld refinement results of powder X-ray diffraction data collected at 300 K and at 85 K (sample made by synthesis route D). Good fits to the reported structural models are observed. The inset of Figure \ref{fig:pxrd}a shows the high temperature tetragonal structure viewed along the (1 1 0) direction. The layered nature of the structure emphasized in Figure \ref{fig:pxrd}a persists into the low temperature orthorhombic structure. The inset in Figure \ref{fig:pxrd}b shows a slab containing one layer of each atom type viewed perpendicular to the layers (along the c-axis). Upon cooling through the phase transition the square nets of atoms present in the tetragonal structure distort into rectangular nets of Fe and of O, and into centered rectangular nets of La and of As. The two Fe-Fe distances in the low temperature phase are labeled in Figure \ref{fig:pxrd}b. For comparison, the Fe-Fe distance in the square nets at 300 K is 2.853 {\AA}.

The structures have been described in detail elsewhere \cite{pub-delaCruz, Nomura}. However, some confusion exists regarding the symmetry of the low temperature structure. It has been reported in both primitive monoclinic and c-centered orthorhombic space groups. In the monoclinic description the reported lattice is metrically orthorhombic and atoms are very close to positions which would give orthorhombic symmetry to the crystal as well. The orthorhombic description is probably correct, and no indication of deviation from orthorhombic symmetry is observed in our PXRD data.

\begin{figure}
\includegraphics[width=3.0in]{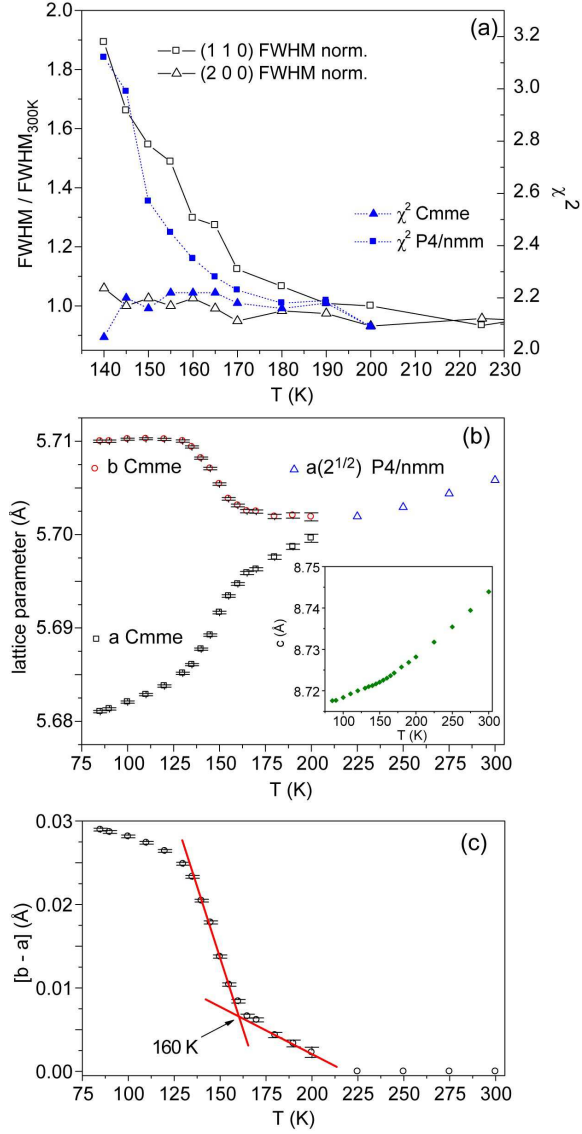}
\caption{\label{fig:FWHM-cells}(color online) Results of temperature dependent PXRD analysis of LaFeAsO showing the continuous nature of the crystallographic phase transition. (a) Solid symbols show $\chi^{2}$ from Rietveld refinement using the orthorhombic low temperature model and tetragonal high temperature model. Open symbols show the normalized width of two Bragg peaks labeled by their tetragonal indices. The (1 1 0) peak is expected to be split into two peaks by the orthorhombic distortion, while the (2 0 0) is not. (b) Unit cell parameters determined by Rietveld refinement. For comparison the \textit{a} lattice parameter in the tetragonal structure has been multiplied by $\sqrt{2}$. Error bars are not shown on some data points for which the error is smaller than the size of the data point symbol. (c) The difference between \textit{b} and \textit{a} in the orthorhombic model, linearly coupled to the order parameter for the structural phase transition. We define the transition temperature for the tetragonal to orthorhombic transition as the kink near 160 K, but note that the structure is evolving continuously both above and below this temperature.
}
\end{figure}

To identify the temperature at which the tetragonal-orthorhombic structural transition occurs, temperature dependent diffraction data were refined in both models. Figure \ref{fig:FWHM-cells}a compares the goodness of fit ($\chi^2$) obtained from each refinement, and shows that the orthorhombic model gives a better fit at 180 K and below. This suggests that the structural transition begins at temperatures significantly higher than previously reported. In addition, we compare in Figure \ref{fig:FWHM-cells}a the width of two Bragg peaks, one which is split by the structural distortion (1 1 0) and one which is not (2 0 0). We note that indices refer to the tetragonal structure. The results show that while the (2 0 0) peak width is unchanged upon cooling, the (1 1 0) peak begins to broaden significantly at about 180 K.

Figure \ref{fig:FWHM-cells}b shows the temperature dependence of the refined lattice constants. Determining the temperature at which to switch between the Cmme and P4/nmm models is not straightforward. We chose to show data in the Cmme model when the difference between the refined values of the \textit{a} and \textit{b} axes is greater than the error associated with these parameters. Based on this criterion, the Cmme model was used at 200 K and below. Above this temperature refinement in this model is in fact unstable and P4/nmm was used. Inspection of the refined values of the lattice constants \textit{a} and \textit{b} reveals upon cooling a gradual divergence below 200 K followed by a more rapid divergence at lower temperatures. The temperature dependence of the lattice constant \textit{c} is shown in the inset of Figure \ref{fig:FWHM-cells}b, and undergoes a change in slope through the structural transition.

The above observations suggest that the structural distortion in LaFeAsO occurs over a wide temperature range, but includes a ``sharp''anomaly as well. To derive a transition temperature T$_{T-O}$ from the structural data, we plot the difference between the lattice constants a and b in Figure \ref{fig:FWHM-cells}c.  These data show an abrupt slope change at about 160 K which we identify as T$_{T-O}$. We note however that the onset temperature of this distortion (180$-$200 K) is significantly higher than T$_{T-O}$. We believe this extension of the lattice distortion to temperatures well above the ``transition temperature'' is observable in the data shown in previous reports which interpreted the results as a single sharp transition \cite{pub-delaCruz, Nomura}. Perhaps the two-dimensional nature of the crystal and electronic structures leads to enhanced fluctuations above the transition temperature.

\subsection{Neutron diffraction: magnetic phase transition}

\begin{figure}
\includegraphics[width=3.0in]{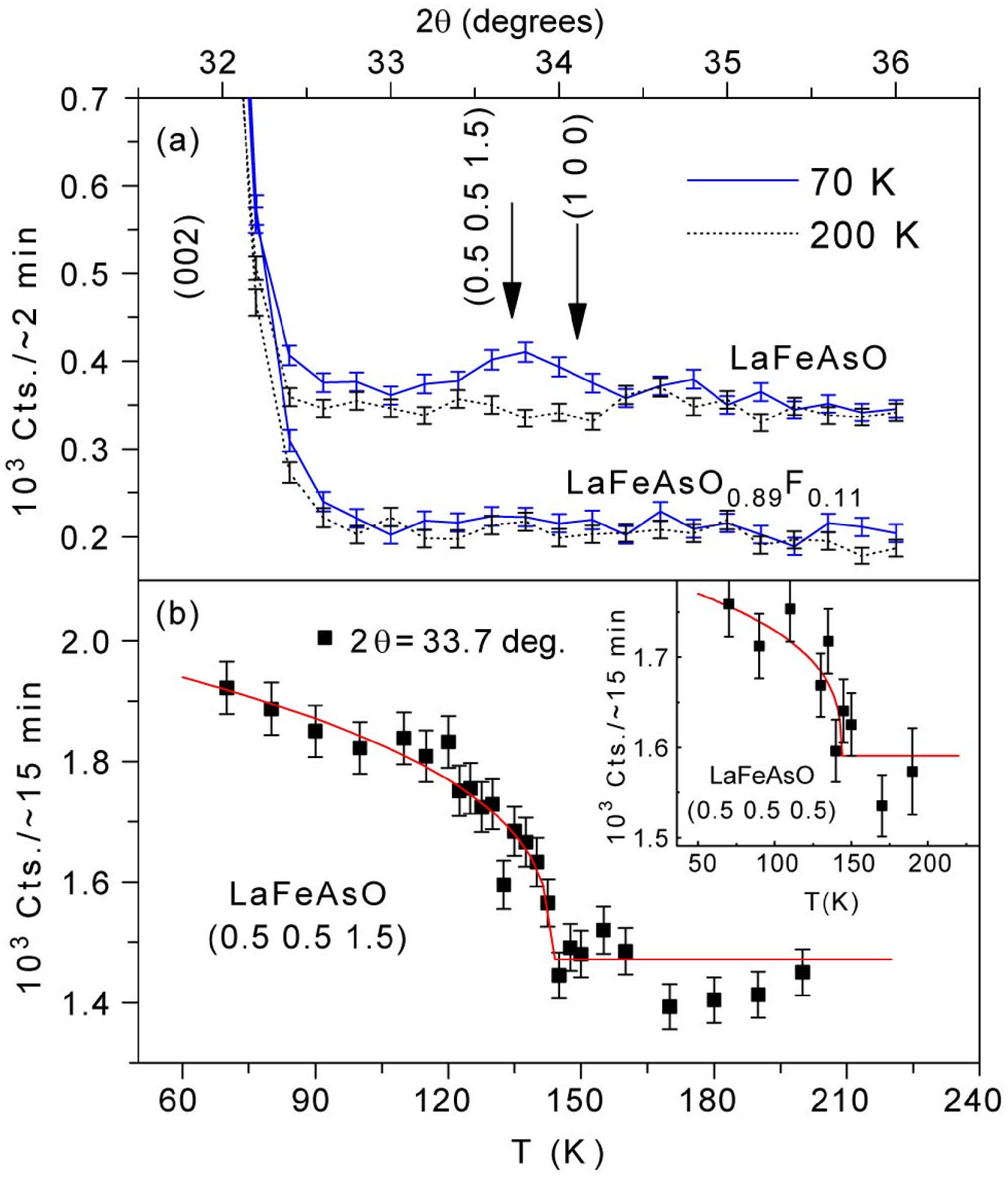}
\caption{\label{fig:neutron}
(color online) (a) Neutron diffraction data for LaFeAsO and LaFeAsO$_{0.89}$F$_{0.11}$ at 70 and 200 K.  Arrows indicate the angular position of the (1 0 0) and (0.5 0.5 1.5) wavevectors.  For clarity, the data for LaFeAsO has been displaced by 150 counts. (b) The temperature dependence of the (0.5 0.5 1.5) and (0.5 0.5 0.5) (inset) positions. The wavevectors are labeled using the tetragonal setting. The lines are a guide to the eye.
}
\end{figure}

Neutron diffraction results reveal a magnetic transition somewhat below the temperature at which the structural transition occurs. Figure \ref{fig:neutron}a shows neutron diffraction data for both LaFeAsO (synthesis route A) and superconducting LaFeAsO$_{0.89}$F$_{0.11}$ \cite{pub-Sefat} at 70 and 200 K from 32 to 36$^{\circ}$ in scattering angle.  These data indicate the presence of additional scattering at 33.7$^{\circ}$ at 70 K that is not present at 200 K.  Figure \ref{fig:neutron}b shows the temperature dependence of the scattering at 33.7$^{\circ}$.  These data show that the additional low temperature intensity in LaFeAsO appears below 145 K. Polarized measurements (not shown) indicate that this scattering is magnetic in nature. The same angular range was explored in LaFeAsO$_{0.89}$F$_{0.11}$. Although the counting time was doubled to 4 minutes per point (normalized in Fig. \ref{fig:neutron}a to 2 minutes per point for the purposes of comparison) there is no discernable difference between the data at 70 and 200 K.  Full refinements of the neutron data indicate similar impurity phases in both samples and, hence, the extra intensity is intrinsic to LaFeAsO. For reference, the strong rise in intensity below 32.5$^{\circ}$ corresponds to the (0 0 2) structural Bragg peak.

The arrows in Fig. \ref{fig:neutron}(a) show the position of two plausible wavevectors (labeled in the tetragonal setting) to describe the additional intensity at 33.7$^{\circ}$ in LaFeAsO.  The position of (1 0 0) is significantly far from the observed peak in the intensity to rule it out as the wavevector. However, the data is strongly consistent with a wavevector of (0.5 0.5 1.5).  To check the indexing of this peak, we searched for extra intensity at (0.5 0.5 0.5).  The intensity observed at this location is very weak but careful measurement of the temperature dependence shown in the inset of Fig. \ref{fig:neutron}(b) indicates an increase below 145 K in a manner consistent with the 33.7$^{\circ}$ reflection.  The observation of these two reflections provides strong evidence that the true ordering wavevector in LaFeAsO is (0.5 0.5 0.5). This indicates a doubling of the conventional unit cell along both a-axes and along the c-axis, and is also consistent with the $\sqrt{2}\times\sqrt{2}$ unit cell suggested by Dong \textit{et al.} \cite{DongArXiv}, and is in good agreement with the magnetic structure derived from other recent neutron diffraction experiments on LaFeAsO \cite{pub-delaCruz}. Evidence of this antiferromagnetic ordering is also observed in the magnetization data discussed below.

\subsection{Elastic constants}

\begin{figure}
\includegraphics[width=3.0in]{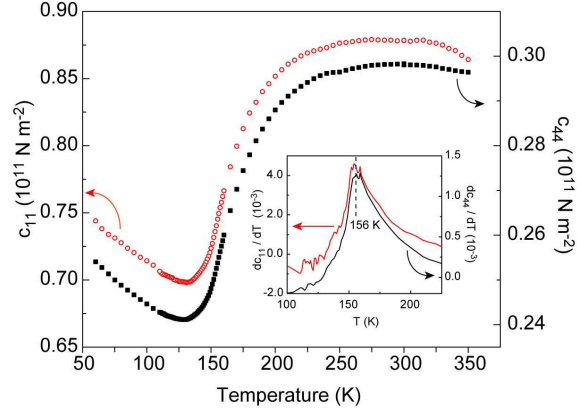}
\caption{\label{RUS}
(color online) Results of resonant ultrasound spectroscopy studies showing the effects of the crystallographic phase transition on the polycrystalline elastic moduli c$_{11}$ and c$_{44}$ of LaFeAsO. A gradual softening begins above 200 K, well above the structural transition, and extends down to 140 K. A relatively sharp peak in the derivative (inset) occurs near T$_{T-O}$.
}
\end{figure}

Resonant ultrasound spectroscopy (RUS) measurements were performed to investigate the elastic properties of LaFeAsO. This technique can be a very sensitive and powerful probe for studying phase transitions, in particular structural transitions in which strain is the order parameter. Figure \ref{RUS} shows the temperature dependence of the polycrystalline elastic moduli c$_{11}$ and c$_{44}$. A remarkably gradual softening is seen to extend up to above 200 K. Data are shown for a sample made by synthesis route B; similar behavior was observed in material from synthesis route A. The temperature derivative of the elastic moduli are shown in the inset of Figure \ref{RUS}. The sharp cusp in dc$_{ij}$/dT occurs at 156 K and is identified as the structural transition temperature determined by this experimental probe. This agrees well with T$_{T-O}$ determined by the structural analysis presented above. The extension of the elastic softening to well above T$_{T-O}$ is consistent with the PXRD data presented above and is further evidence of the gradual nature of the crystallographic phase change in this material. We also note that no significant thermal hysteresis or magnetic field effect on the elastic properties was observed.

\subsection{Heat capacity}

\begin{figure}
\includegraphics[width=3.0in]{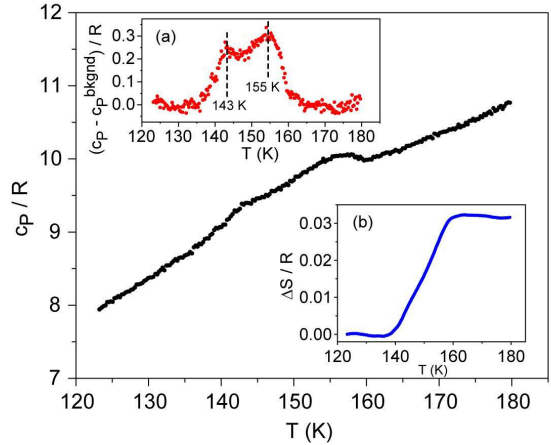}
\caption{\label{cP}
(color online) The measured heat capacity of LaFeAsO (in units of R per formula unit) which displays broad anomalies at both phase transitions. Inset (a) shows the baseline subtracted data which clearly shows two broad peaks centered near the magnetic and structural transition temperatures. Inset (b) shows the integrated entropy associated with the two anomalies.
}
\end{figure}

The measured heat capacity of LaFeAsO (synthesis route A) in the vicinity of the structural and magnetic transitions is shown in Figure \ref{cP}. Two broad overlapping anomalies are observed. We note that two heat capacity anomalies were also observed in sample used for PXRD analysis, produced by route D. Baseline subtracted data are shown in Figure \ref{cP} inset (a). The background was estimated by a polynomial fit to the data above and below the transition region. The subtracted data clearly show two peaks, one associated with the structural transition centered at 155 K, and one due to the magnetic transition centered at 143 K. The coincidence of these two peaks with the structural and magnetic phase transition temperatures strongly suggest that these are indeed separate anomalies, and not a single transition smeared by, for example, inhomogeneities. No hysteresis was observed in repeated measurements, suggesting that the phase transitions are second order or perhaps weakly first order. The entropy change determined by integration of the subtracted heat capacity data is shown in Figure \ref{cP} inset (b). A total entropy of 0.032 R (0.27 J K$^{-1}$ mol FU$^{-1}$) is determined by the integration. If the total change in entropy $\Delta$S is considered purely electronic ($\Delta S = T_C\Delta\gamma$) then $\Delta\gamma \sim $1.7 mJ mol$^{-1}$ K$^{-2}$.

\subsection{Transport properties}

\begin{figure}
\includegraphics[width=3.0in]{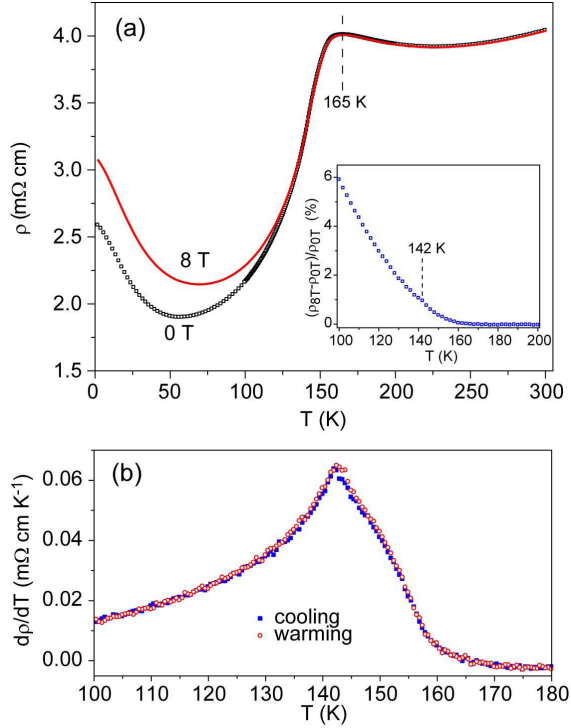}
\caption{\label{rho}
(color online) Effects of the phase transitions on the electrical transport of LaFeAsO. (a) The temperature dependence of the electrical resistivity with no applied magnetic field and with an applied field of 8 T. The inset in (a) shows the magnetoresistance calculated from the resistivity data. (b) The temperature derivative of the measured resistivity on cooling and warming illustrating the absence of thermal hysteresis. The effect of the structural transition at 160 K is shown. The peak in d$\rho$/dT is near the magnetic transition temperature at 143 K.
}
\end{figure}

The transport properties reported here were all measured on a single sample produced using synthesis route A. Figure \ref{rho}a shows the measured electrical resistivity of LaFeAsO, which agrees well with previous reports \cite{KlaussArXiv}. At room temperature $\rho$ has a value of 4 $m\Omega cm$ and decreases upon cooling. This is typical of a low carrier concentration metal or heavily doped semiconductor. The upturn in $\rho$ on cooling below about 200 K is likely due to increased charge carrier scattering by lattice fluctuations related to the onset of the impending structural transition. The electrical resistivity reaches a local maximum at 165 K and drops rapidly below this temperature.

The temperature derivative of the measured resistivity data is shown in Figure \ref{rho}b. Below about 160 K a sharp increase in d$\rho$ /dT is observed. The maximum occurs at 143 K, near the magnetic transition temperature.  No thermal hysteresis is observed in these data, suggesting that neither of the phase transitions are strongly first order. Since the resistivity anomaly begins near T$_{T-O}$ = 160 K and this is the temperature below which significant magnetoresistance appears (Figure \ref{rho}a inset), we believe that structural phase transition in LaFeAsO is primarily responsible for the dramatic changes in transport properties. The magnetic transition occurs near the peak in d$\rho$/dT, and may be responsible for the weak anomaly observed at 142 K in the magnetoresistance (Figure \ref{rho}a inset).

\begin{figure}
\includegraphics[width=3.0in]{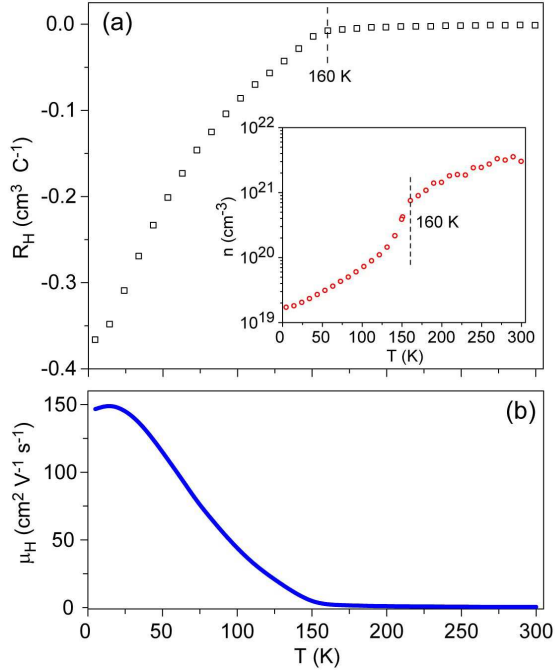}
\caption{\label{Hall}
(color online) Results of Hall effect measurements on LaFeAsO showing the remarkable decrease in inferred carrier concentration and high mobility at low temperatures. (a) The measured Hall coefficient and inferred carrier concentration (inset). (b) The Hall mobility calculated from the carrier concentration in (a) and the electrical resistivity (Figure \ref{rho}).
}
\end{figure}

Results of Hall effect measurements are shown in Figure \ref{Hall}. Near room temperature the Hall coefficient is negative and nearly temperature independent, indicating conduction by electrons with an inferred concentration n = $3\times10^{21} cm^{-3}$. We note that the presence of multiple bands at the Fermi level complicates the interpretation of the Hall coefficient data. In a simple model with one electron band and one hole band the inferred value of n gives an upper bound on the concentration of the dominant charge carriers. Upon cooling below 160 K the Hall coefficient decreases rapidly, and \textit{n} drops by an order of magnitude between 160 and 100 K (Figure \ref{Hall}a, inset). This suggests that many of the charge carriers present in the high temperature phase are localized at T$_{T-O}$ due to the structural transition. This is possibly related to the local moment formation and subsequent magnetic ordering observed below T$_{T-O}$. It is interesting that this order of magnitude decrease in n coincides with a factor of two decrease in $\rho$. This suggests a large change in carrier mobility. The Hall mobility $\mu_{H} = R_H\rho^{-1}$ is plotted in Figure \ref{Hall}b. The mobility increases by a factor of 300 between room temperature and 5 K, and reaches the remarkably high value of 150 $cm^2V^{-1}s^{-1}$ in this polycrystalline material.

\begin{figure}
\includegraphics[width=3.0in]{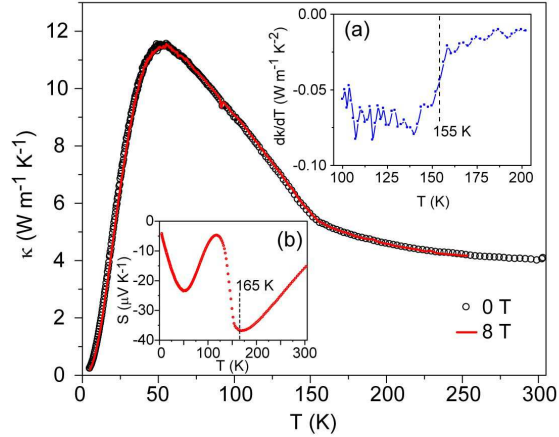}
\caption{\label{Sk}
(color online) Thermal and thermoelectric transport properties of LaFeAsO. The thermal conductivity measured in zero applied magnetic field and in a field of 8 T, showing an abrupt change in slope (inset a) as the structure transforms from tetragonal to orthorhombic upon cooling. Inset (b) shows the dramatic changes in the Seebeck coefficient that occur below the transition temperatures.
}
\end{figure}

The measured thermal conductivity $\kappa$ and Seebeck coefficient S of LaFeAsO are shown in Fig. \ref{Sk}. The thermal conductivity increases abruptly below about 155 K, but otherwise follows the behavior of typical crystalline materials. The increase in $\kappa$ cannot be attributed to the change in electronic thermal conduction as the thermal conductivity in this low carrier concentration material is dominated by phonons. Thus the observed behavior must be attributed to an increase in the thermal conductivity of the lattice, and related to the tetragonal-orthorhombic crystallographic transition that occurs at 160 K (Figure \ref{Sk} inset a). This could be due to a decrease in electron-phonon scattering below the transition, which would suggest strong coupling between the charge carriers and the lattice vibrations in the tetragonal phase of LaFeAsO through bond length fluctuations. This is also consistent with the rapid increase in the carrier mobility presented above (Figure \ref{Hall}b). An increase in phonon thermal conductivity could also arise from the freezing in of phonon-scattering lattice fluctuations upon cooling through T$_{T-O}$.

The Seebeck coefficient (Fig. \ref{Sk} inset b) is negative over the entire temperature range, indicating that electrons dominate the electrical conduction. This is consistent with the negative Hall coefficient (Figure \ref{Hall}). The Seebeck coefficient is moderately high in this material, but lower by about a factor of two than in superconducting LaFeAsO$_{0.89}$F$_{0.11}$ above $T_{C}$ \cite{pub-Sefat}. A remarkable decrease in $|S|$ is observed below about 155 K. It is unusual to see a sharp drop in $|S|$ coinciding with a sharp drop in carrier concentration. Boltzmann transport theory predicts for the free electron model \cite{McCarten1994}
\begin{equation} \label{eq:S}
S(T) = -\frac{\pi^2}{3} \frac{k_B}{|e|} k_{B} T \left[\frac{N(\epsilon_F)}{n}+\frac{1}{\tau(\epsilon_F)}
\frac{d\tau}{d\epsilon_F}\right].
\end{equation}
The observed decrease in both $|S|$ and \textit{n} through the transition indicates that the second term in Eq. \ref{eq:S} is dominant in this temperature regime. This suggests that the charge carrier scattering mechanism is changed significantly as the material passes through the phase transition region, and presents further evidence for the reduction of electron-phonon interactions in the orthorhombic phase which was suggested by the above analysis of carrier mobility and thermal conductivity. This is evidence of strong electron-phonon coupling in LaFeAsO.

\subsection{Magnetic properties}

\begin{figure}
\includegraphics[width=3.0in]{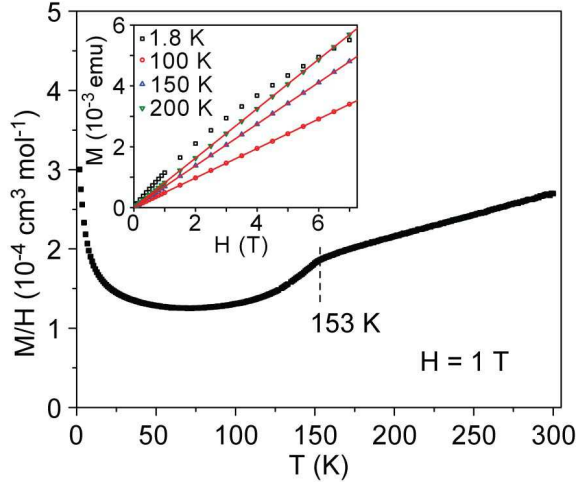}
\caption{\label{mag}
(color online) The temperature dependence of M/H (per mole of formula units) for a PXRD pure polycrystalline LaFeAsO sample. The inset shows M(H) data at four different temperatures.
}
\end{figure}

Results of magnetization measurements on a PXRD pure polycrystalline sample of LaFeAsO (synthesis route C) are shown in Figure \ref{mag}. Measured M(H) curves (Fig. \ref{mag} inset) show a paramagnetic response up to H = 6 T at 100 K, 150 K, and 200 K. The upturn in M/H at low temperatures and the nonlinear behavior of M(H) at 1.8 K are likely due to small amounts of impurities. Other than the Curie tail at low temperatures, the magnetization shows little temperature dependence above and below the transition region. A decrease in M/H is observed through the phase transitions, beginning near 153 K. This has been identified as a signature of spin density wave formation. Localization of charge carriers which occurs at T$_{T-O}$ would also result in a decrease in M/H through the reduction of the Pauli paramagnetism; however, one would expect Curie-Weiss behavior below this temperature if that were the complete story. Neutron diffraction results presented above and elsewhere \cite{pub-delaCruz} suggest an antiferromagnetic ordering develops near the temperature at which M/H decreases in this material. Carrier localization followed closely by antiferromagnetic ordering could be responsible for the behavior observed near 153 K in Fig. \ref{mag}. The magnetic behavior of this anisotropic material is certainly complex and not yet well characterized or well understood. Future studies of single crystals with well controlled stoichiometry will be of great importance to the understanding of the magnetic nature of LaFeAsO.

\begin{figure}
\begin{center}
\includegraphics[width=3.0in]{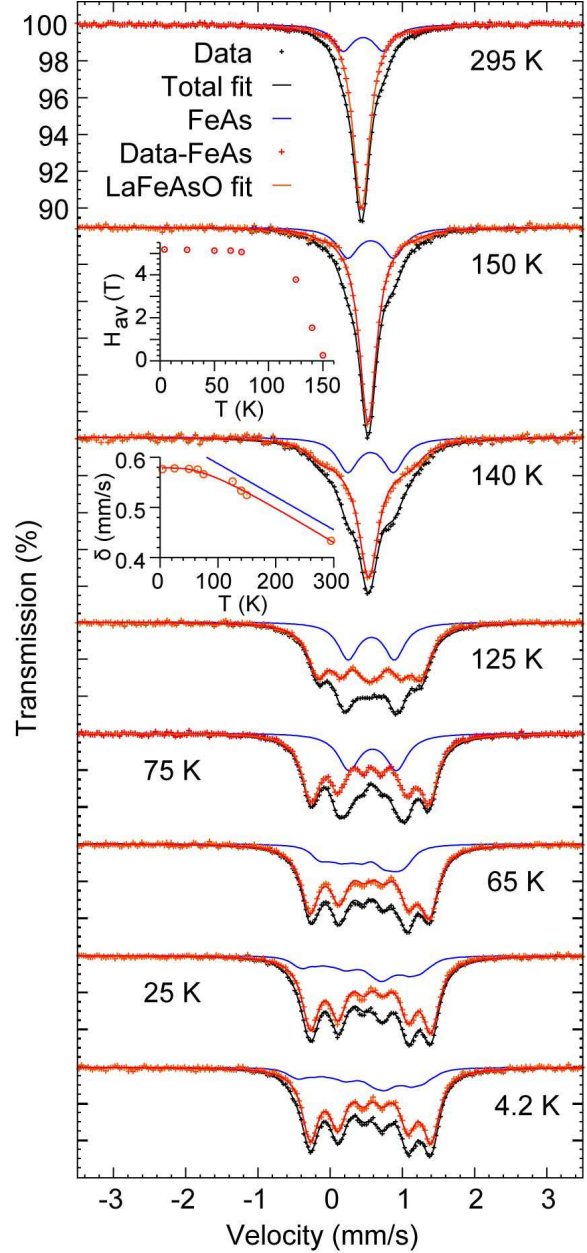}
\end{center}
\caption{\label{Moss}(color online) The M\"ossbauer spectra of LaFeAsO. The total
data and fits, as well as the data with the FeAs impurity contribution
substracted and the LaFeAsO phase fit are shown. Upper inset: the
average hyperfine field determined in the LaFeAsO phase. Lower inset:
The temperature dependence of the isomer shift of LaFeAsO, with a Debye
model fit, and, straight line, the isomer shift in FeAs from Ref. \cite{BB}.
}
\end{figure}

The M\"ossbauer spectra of LaFeAsO produced by synthesis route B are shown in Fig. \ref{Moss}. The spectra were modeled using a mixture of LaFeAsO and FeAs (\textit{vide infra}). Between 4.2 and 75 K, the LaFeAsO M\"ossbauer spectrum is a simple magnetic sextet. At 4.2 K, the isomer shift and quadrupole shift of LaFeAsO are 0.576(5) and -0.031(1) mm/s, respectively, indicating the low-spin nature of the iron(II). The hyperfine field is 5.19(1) T and the usual conversion factor of 15 T per 1 $\mu_B$ yields an estimated Fe moment of ca. 0.35 $\mu_B$. These hyperfine parameters are in very good agreement with those reported in Ref. \cite{KitaoArXiv, raffius1993, AA}. A recent theoretical study has also derived 0.25-0.35 $\mu_B$ in the low temperature phase \cite{Castro-Neto}. Note that in Ref. \cite{raffius1993} the LaFeAs sample was oxidized and that the spectra are actually those of LaFeAsO. The small, but non-zero, quadrupole shift indicates that there is a small lattice contribution to the iron(II) quadrupole interaction, as expected for low-spin iron(II) in a slightly distorted tetrahedral environment. The observed quadrupole shift is constant up to 140 K. At 150 and 295 K the quadrupole splitting is zero within the error bar. The introduction of a small, temperature independent, texture for the LaFeAsO phase, with intensity ratios of 3:2.3:1:1:2.3:3 yielded a significant reduction in $\chi^2$ and this texture was used for the fits in Fig. \ref{Moss}. The temperature dependence of the isomer shift is in very good agreement with the Debye model for the second order Doppler shift and yielded a M\"ossbauer temperature of 294(20) K, see lower inset in Fig. \ref{Moss}. The magnetic moment is essentially constant up to 75 K, and the gradual reduction of the hyperfine field above 125 K indicates a transition to a paramagnetic phase close 150 K. Between 125 and 150 K, the LaFeAsO spectrum is more complex, as in Refs. \cite{KitaoArXiv, KlaussArXiv}, and was modeled herein as a superposition of a sextet and a singlet, and the weighted average hyperfine field is shown in the upper inset of Fig. \ref{Moss}. The observed spectra between 125 and 150 K could be explained either by a small compositional variation and a smearing of $T_N$ or by a hyperfine field distribution resulting from a incommensurate or commensurate spin density wave \cite{KitaoArXiv, KlaussArXiv} or from spin glass behavior in this temperature range. At 75 K and below, however, the simple sextet spectrum is indicative of no such spin density wave or spin glass behavior. Finally, we observe a marginal increase in linewidth on cooling from 295 to 150 K, this increase could be related to the tetragonal to orthorhombic structural distortion.

Because the M\"ossbauer spectra of LaFeAsO and FeAs \cite{haggstrom, BB} are located in the same velocity region, detecting FeAs impurities requires careful comparison of the spectra near the 77 K Neel temperature of FeAs. It is clear that the transition of FeAs from the helimagnetic to paramagnetic phase is responsible for the modification of the spectra between 65 and 75 K. The FeAs contribution above and below 70 K was found to be consistent with and then constrained to the hyperfine parameters for the corresponding temperatures of Ref. \cite{BB} and Ref. \cite{haggstrom}, respectively, and resulting fits are of very good quality. The FeAs contribution corresponds to ca. 10 \% weight of the sample, a quantity that is essentially temperature independent, indicating a similar temperature dependence of the recoil free fraction in FeAs and LaFeAsO and indeed the ca. 285 K M\"ossbauer temperature obtained from the second order Doppler shift in FeAs, Ref. \cite{BB} and the line in the lower inset in Fig. \ref{Moss}, is very close to 294 K observed herein for LaFeAsO. The only free fit parameters for the FeAs phase are the linewidth, and the isomer shift and average hyperfine field at 25, 50, and 65 K. We believe that the larger intensity of the 2:5 lines in the spectra of Refs. \cite{AA, KlaussArXiv, KitaoArXiv} arises not from a possible texturing of the sample but from an underlying FeAs impurity subspectrum. The presence of this FeAs impurity, if not properly modeled will lead to a smaller effective hyperfine field, as seen in Ref. \cite{KlaussArXiv}. Further, it is likely that the third component observed in $\mu$SR measurements below 70 K \cite{KlaussArXiv} is also related to the FeAs impurity that is clearly visible in the 78 K M\"ossbauer spectrum on the same sample.

\section{Conclusions}

At high temperatures LaFeAsO is a low carrier concentration metal with conduction dominated by electrons and with no local magnetic moment. This is supported by measurements of electrical resistivity (Figure \ref{rho}), Hall coefficient and carrier concentration (Figure \ref{Hall}), Seebeck coefficient (Figure \ref{Sk}), and magnetization (Figure \ref{mag}). Upon cooling a crystallographic phase transition occurs. We propose that the structural transformation occurs continuously below about 200 K based on structure refinements (Figure \ref{fig:FWHM-cells}) and elastic response (Figure \ref{RUS}), and that the kink in the order parameter b-a at 160 K (Figure \ref{fig:FWHM-cells}c) indicates the point at which sufficient distortion has occurred to result in carrier localization and local moment formation on the Fe atoms. This is supported by the small entropy associated with the transition at T$_{T-O}$ (Figure \ref{cP}). It is unclear what drives this structural transition. One candidate is a band Jahn-Teller effect \cite{BandJT}, in which energy is gained by splitting the sharp peak in the density of states near the Fermi level \cite{SinghArXiv} by distorting the crystal structure. This could result in localization of some conduction electrons and the development of a local magnetic moment on Fe. Careful theoretical analysis of the effects of structural distortion on the band structure of LaFeAsO could help resolve this issue. Upon further cooling the local moments order near 145 K (Figure \ref{fig:neutron}). Strong electron-phonon coupling exists in the high temperature tetragonal phase, as evidenced by the behavior of the mobility (Figure \ref{Hall}b), thermal conductivity, and Seebeck coefficient (Figure \ref{Sk}) through the phase transition region. Upon doping with, for example, fluorine, the structural phase transformation and associated magnetic ordering is suppressed (Figure \ref{fig:neutron}), allowing the strong electron-phonon coupling present in the tetragonal phase to extend down to low temperature, and superconductivity emerges. It is believed that the phase transition is suppressed electronically by electron or hole doping with substitutional F or Sr, or with oxygen vacancies. However, disorder may also suppress the structural transition. Perhaps new superconductors can be discovered by investigating isoelectronic substitutions which suppress T$_{T-O}$ in the ZrCuSiAs-type rare-earth iron oxyarsenides and related ThCr$_2$Si$_{2}$-type alkaline-earth iron arsenides.

We acknowledge useful discussions with D. J. Singh, O. Garlea, S. Nagler, A. Castro-Neto, and I. I. Mazin, and technical assistance by J. Zarestky and J. Fernandez-Baca. Research sponsored by the Division of Materials Sciences and Engineering, Office of Basic Energy Sciences.
Part of this research performed by Eugene P. Wigner Fellows at ORNL, managed by UT-Battelle, LLC, for the U.S. DOE under Contract DE-AC05-00OR22725.
A portion of this research at ORNL's Center for Nanophase Materials Sciences and High Flux Isotope Reactor was sponsored by the Scientific User FacilitiesDivision, Office of Basic Energy Sciences, DOE.
A portion of this work was supported by NSF grant DMR-0400938.
F.G. acknowledges with thanks the financial support of the Fonds National de la Recherche Scientifique, Belgium, through grants 9.456595 and 1.5.064.05.


\end{document}